\documentclass[aps,prl,twocolumn,groupedaddress,showpacs,amsmath,amssymb]{revtex4}
\usepackage{epsfig}
\usepackage{amssymb}
\usepackage{amsmath}
\usepackage{graphicx}
\usepackage{amsfonts}
\usepackage{epsfig}
\usepackage{revsymb}
\usepackage{bm}
\usepackage{latexsym}
\usepackage{hyperref}

\begin{document}

\title{Quantum crystal oscillations in a superfluid liquid}

\author{V. L. Tsymbalenko}
\email[]{vlt@isssph.kiae.ru} \affiliation{Kurchatov Institute, 123182 Moscow, Russia}


\begin{abstract}
We study the hydrodynamics of quantum $^4$He crystal oscillations in a superfluid liquid
with involving the dynamics of atomically rough surfaces. It is shown that,
due to enhancement of the kinetic growth coefficient as the temperature lowers,
the reaction force of the liquid applied to the $^4$He crystal changes its character
from the inertial to viscous one vanishing as $T \rightarrow 0$.
The model is confirmed by the experiments on the oscillations of the $^4$He crystal
within the temperature range 0.54-1.43~K at frequencies 484 and 211~Hz.
A new type of hydrodynamic instability is found.
The instability occurs provided that the oscillation amplitude of the velocity
becomes higher than $\sim 3 ~cm / sec$.
\end{abstract}


\maketitle

\section{INTRODUCTION}
\par
The motion of a helium crystal in superfluid $^4$He liquid differs from the motion of the conventional solids in their
 liquids. The distinction, in the first turn, is associated with the high growth kinetics of the phase liquid-crystal 4He
 interface \cite{ABP}. In fact, the crystallization occurs in the regions with the pressure above the phase equilibrium
 pressure. On the other hand, the melting takes place in the regions with the pressure below the phase equilibrium
 pressure. As a result, the shape of a crystal and the fluid flow around the crystal can vary as the pressure changes
 beside the crystal. In addition, the mass flow across the interface appears and changes the boundary conditions
and, finally, force applied to the crystal. The kinetic growth
coefficient $K$ depends on the temperature and varies by several
orders of the magnitude within the range 0.4-1.4K. This gives a
unique possibility to study experimentally the effect of phase
transition dynamics on the hydrodynamics of the motion in a
superfluid liquid from the case of almost immobile crystal surface
at high temperatures to the highly mobile surface below
$\sim$0.55K. Another problem that could be solved by setting the
crystal into the oscillatory motion is associated with the
"burst-like growth" effect \cite{VLT2,Finn}. The excess pressure
at certain temperatures increases
 the growth rate of
the crystal facets by several orders of the magnitude. This fast-growing state continues to exist for several
milliseconds after relaxing the pressure to the phase equilibrium pressure. It is possible that the periodic
 dynamic pressure of the fluid induces such transition and stabilizes the anomalous state.

\section{SURFACE CONTRIBUTIONS TO THE DYNAMIC RESPONSE OF  CRYSTAL}
\par
Confining ourselves with the experimental conditions, we consider the temperature range 0.4-1.42~K,
the frequencies $\sim$200 and $\sim$500~Hz and velocities of the crystal motion much less than the speed of sound.
In this case, the superfluid hydrodynamics can be described by two independent equations
for the superfluid (s) and the normal (n) components~\cite{LL}. In addition, the conditions
of small oscillations are fulfilled in the experiment. The oscillation amplitude $A$
is much smaller than the size of the crystal $R$.
This allows us to restrict ourselves with the linear terms in the equations of motion.
  In fact, the pressure due to acceleration motion is about $\delta p_{acc}  \sim \rho R a \sim \rho R A \omega^2$,
 and the Bernoulli pressure is $\delta p_B \sim \rho v^2 \sim \rho A^2\omega^2$.
 For $A \ll R$, the condition $\delta p_{acc}  \gg \delta p_B$ holds for.
\par
Let us consider the case of a spherical crystal with radius $R$,
set into oscillations with frequency $\omega$ and rate $u = \omega A$.
The superfluid motion obeys the equations of an ideal fluid
\begin{equation}\label{f01}
\overrightarrow v_s=\nabla \varphi_s,\;\;    \triangle \varphi_s = 0,\;\;  p_s=-\rho_s \frac{\partial \varphi_s}{\partial t}\, ,
\end{equation}
For the temperatures when $\rho_n \ll \rho \approx \rho_s$, the situation corresponds to the streamline of the sphere with an ideal fluid, $v \approx v_s$  and $p \approx p_s$ . The boundary condition reads
\begin{equation}\label{f02}
\overrightarrow v=\overrightarrow{u} -\overrightarrow{V}\frac{\Delta \rho}{\rho},\;\;
 V=K \delta \mu = K \frac{\Delta \rho}{\rho' \rho}p,\;\;  \Delta \rho = \rho' - \rho\,,
\end{equation}
where $\rho'$ is the density of the solid phase, $V$ is the normal growth rate of the surface, and $K$ is the growth
 kinetic coefficient. Solving equations Eqs.~(\ref{f01}) under conditions (\ref{f02}),
 we obtain a stationary solution for harmonic $l~=~1$. The force applied to the crystal is given by expression
\begin{equation}\label{f03}
F_s = \frac{2 \pi}{3} \rho R^3 \frac{i \omega u}{1+i \omega R \frac{K}{2}\frac{\Delta \rho^2}{\rho \rho'}}.
\end{equation}
For high temperatures, coefficient $K$ is small and we obtain the usual expression for the associated mass.
For low temperatures and $K \rightarrow \infty$, the force of reaction has a viscous nature and vanishes.
In this limit the movement of the crystal surface is practically compensated by its remelting, so that the surface is nearly motionless. The harmonics with $l \geq 2$ have the following spectrum~\cite{BDT}
 \begin{equation}\label{f04}
 \omega _l^2 - i\omega _l\,\frac{\rho \rho '}{\Delta \rho^2 K}\,\frac{l+1}{R}
 -\alpha \frac{ (l^2-1)(l+2)}{R^3}\frac{\rho}{\Delta\rho^2} =0,
\end{equation}
where $\alpha$ is the surface tension which we assume isotropic for simplicity.
These harmonics cannot by generated by the longitudinal oscillations of the crystal.
As the temperature increases, the associated mass of the superfluid component decreases.
The force of reaction in Eq.~(\ref{f03}) should be multiplied by a factor $\rho_s / \rho$.
\par
The flow of the normal component along the crystal surface contributes to the inertial and the dissipative
parts of the force. The equation for the normal component hydrodynamics reads
 \begin{equation}\label{f05}
\rho_n \frac{\partial \overrightarrow{v}}{\partial t} = - \nabla p_n + \eta \Delta \overrightarrow{v_n}.
\end{equation}
Following Andreev and Knizhnik~\cite{AK}, we put the normal component velocity to be the same
as the interface rate $V$. Thus, the flow of the normal component across the boundary is absent,
and the crystal can be treated as a body with an impermeable surface.
The general solution of Eq.~(\ref{f05}) for the oscillations of a sphere in a viscous fluid is given in Ref.~\cite{LL}.
This expression can be simplified since the penetration depth is $\delta_{vis} = \sqrt{2 \eta / \rho_n \omega} \ll R$ in our experiments.
The viscous flow decays in the narrow layer beside the crystal interface, while the fluid flow is potential in the rest volume. In this limiting case the force acting on the crystal is given by the expression
 \begin{equation}\label{f06}
F_n = \frac{2 \pi}{3} \rho_n R^3 \frac{du}{dt} + 3 \pi R^2 \sqrt{2 \eta \rho_n \omega}u.
\end{equation}
From Eqs.~(\ref{f03}) and (\ref{f06}) it follows that the inertial contribution depends weakly on the temperature.
The decrease of the superfluid component contribution is compensated by increasing the normal component
 contribution. A ratio of the second dissipative term to the first inertial one is about
 $\sim \delta / R \sim 10^{-3} \ll 1$.
 Thus, the total dissipative part of the force is determined mainly by the surface growth kinetics.
\par
The above results are applicable to the temperatures higher than $\sim$1.2K when the shape of the free growing
crystal is nearly spherical. The example of such growth is given in Ref.~\cite{VLT6}. For the further cooling, two
roughening transitions are observed so that the crystal below $\sim$0.9K has a shape of the hexagonal prism.
The facets of the prism are connected with the atomically rough surfaces.
The facets have a slow kinetics and can be treated as immobile in our experiments.
Then the crystallization and melting will occur at the atomically rough segments occupying a small area.
For small oscillations, when the growth amplitudes of these segments are much smaller than their radius of curvature, the estimate of the force gives an expression similar to Eq.~(\ref{f03}). One should only to replace the sphere radius with a typical crystal size $L$ and the coefficient $K$ should be multiplied by factor $\xi$. The latter is about a ratio of the area of the active segments to the total crystal area. For example, the visual estimate of the crystal surface gives us an approximate ratio of the rough segments area to the total area of the crystal $\xi\sim0.1$
Since the real shape of the crystal is far from spherical, our estimates are rather crude and thus have qualitative character sufficient for our purposes.
\par
\begin{figure}
\includegraphics[scale=1]{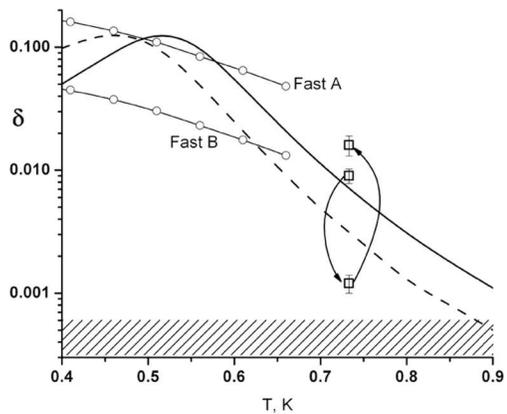}
\caption{Temperature dependence of the oscillator damping. The shaded area is a natural damping of the oscillator. The explanations are in the text.}
\label{fig1}
\end{figure}

The essential effect of melting and crystallization is observed at the temperature for which the condition
 \begin{equation}\label{f07}
\xi K(T^{\ast}) \frac{\Delta \rho ^2}{\rho \rho'} \omega L \sim 1
\end{equation}
holds for. For frequencies $\sim$200 and $\sim$500 Hz, the crystal size $L \sim 0.1 cm$, $\xi = 1$ and the evaluation with Eq.~(\ref{f07}) yields $T^{\ast}$ = 0.62-0.68K. This is an overestimate since
the crystal is faceted at such temperature. Provided the amount of the atomically rough segments
is one tenth of the total area ($\xi$ = 0.1),
the maxima of the damping lie within the range $T ^{\ast}$ = 0.49-0.54K, see Fig.~\ref{fig1}.
\par
For large amplitudes, the regime becomes non-linear since the atomically rough segments are geometrically
cut by the facets. In the region of the oncoming flow the crystallization reduces their area and increases
the surface curvature. This results in reducing the difference between chemical potentials.
As a result, the mass flow across the interface decreases. On the opposite side of the crystal the melting
takes place with increasing the curvature radius of the atomically rough surface segments.
Since the volume is closed, the mass flow into the liquid results in the total increase of the pressure in
the container. This pressure should be added to the dynamic pressure. The parameters of the influence
will depend on the size of the crystal, oscillation amplitude, and the container volume in which the crystal
is grown. The crossover to the nonlinear regime occurs at the oscillation rates as
 \begin{equation}\label{f08}
u^{\ast} \sim \frac{\rho'}{\Delta \rho}\frac{1}{K} \sim \frac{10}{K}.
\end{equation}
For T = 0.5-0.8K, the magnitudes of the critical velocity $u^{*}$ = 2-200 cm / s.

\section{EXPERIMENTAL METHODS}
\par
\begin{figure}
\includegraphics[scale=1]{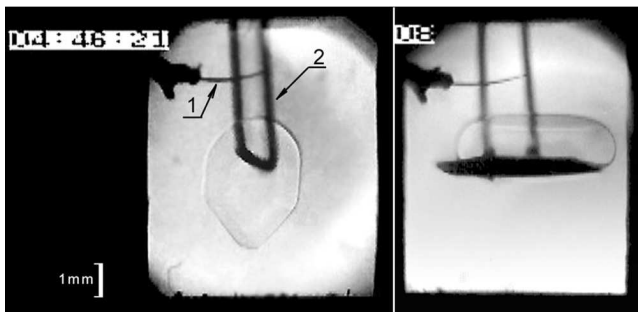}
\caption{Left: the first series, T=0.48~K. The crossbar pierces the crystal. Right: the second series, T=0.84~K. A crystal is located at the platform glued to the loop and (1) tungsten tip, (2) a superconducting loop.}
\label{fig2}
\end{figure}

To set the crystal in motion, we use an oscillator with the superconducting U-shaped loop which the crossbar
 is placed in a constant magnetic field $\sim$3 kOe. A tungsten needle is located above the crossbar.
 The crystal nucleates at the needle with applying high voltage which induces an additional supersaturation
 near the surface of the tip due to electrostatic pressure. Then the crystal remelts down to the crossbar of the
 oscillator. We use two types of oscillators, see Fig.~\ref{fig2}.
 The first represents a superconducting wire with the
 resonance frequency 484~Hz. The second is made with a glassy table glued with epoxy to the crossbar in order
 to reduce the frequency to 211~Hz. The first design is suitable for the measurements below 0.9~K, when
 the crystal is cut by the facets and remains at the crossbar for a long time. The second design of the oscillator (see Fig.2, right) prevents
 from the remelting of the crystal in the hydrostatic pressure gradient and descending the crystal to the bottom
 of the container. The latter design can be used within the whole temperature range.
The oscillations of the loop are generated with an alternating current. The voltage arising due to the motion of a
 crossbar in the magnetic field is amplified and registered with the computer-supplied measurement system.
 We use two methods. In the first one the amplified voltage of the loop is fed to the digital amplitude
 stabilizer\cite{VLT7} and transferred after the phase shifter and current amplifier back to the loop.
 The self-exciting oscillator allows us to measure continuously the frequency and the damping of the system
 at fixed oscillation amplitude. The absolute magnitudes of the decrement and frequency are determined
 from the free damped oscillations after the break-up of feedback. The second method represents
 a measurement of the amplitude-frequency characteristics of the system.
 This method is employed for large crystals yielding high attenuation and  frequency shift into the system.
\par
In the experiment we use the oscillation mode corresponding to the synchronous oscillations of the vertical parts of the loop. The horizontal crossbar moves as a whole, producing an additional inertial force at the free ends of the vertical segments of loop. The equation describing this mode is derived from the equation for the vibrations of the rod with one fixed end\cite{LL2}. For derivation we augment the inertial force resulted from the finite mass of a half of the crossbar, exerted to the free end of the vertical segment of the loop. Finally, we arrive at
\begin{gather}\nonumber
\cos(kL)\cosh(kL)+1+\nonumber
\\ +akL[\cos(kL)\sinh(kL)-\sin(kL)\cosh(kL)]=0\nonumber
\\
k^4=\frac{4 \rho_{loop} \omega^2}{r^2 E},\;  \tau=\frac{\xi}{2}K(T)R\frac{\Delta\rho^2}{\rho\rho'},\;\nonumber
\\
 a=\frac{1}{M_{0,1}}(M_1 + M_c +M_c \frac{\rho}{2 \rho'}\frac{1}{1+i \omega \tau}). \label{f09}
\end{gather}
Here $L$ is the length of the vertical segments of the loop, $r$ is the radius of the wire, $E$ is its Young modulus,
 $M_{0,1}$ are the masses of the vertical segments and horizontal crossbar, $M_c$ is the helium crystal mass,
 and $R$ is the radius of the crystal.
 An additional bending stress at the vertical ends, associated with the rotating oscillations of the crossbar, changes slightly the main mode frequency. In general, there exist other oscillation modes, for example, the torsional mode with the out-of-phase motion of vertical parts of the loop or the modes with exciting the proper oscillations of the crossbar. These modes have higher frequencies as compared with the principal mode described by Eq.~(\ref{f09}). In our experiment these additional modes were not excited.
\par
The damping rate is determined by imaginary part of the frequency $\omega$ found from solving Eq.~(\ref{f09}). We define the decrement of the system $\delta$ as the logarithm of the damping of the oscillation amplitudes $A$ for the period, $\delta=\ln(A(t)/A(t+T_0))$, $T_0$ being the oscillation period. The decrement is directly connected with the imagenary part of the frequency, $\delta=2\pi\frac{Im(\omega)}{Re(\omega)}$.
\par
Using the data on the temperature behavior of the kinetic growth coefficient, in Fig.~\ref{fig1} we show the calculation for the
 attenuation of the composite oscillator with a crystal of radius 1 mm. Above the first roughening transition
 temperature $\sim$1.25K the decrement of the system at the both frequencies is less than $10^{-3}$.
 This value is smaller  than the damping of the free loop. Below the both roughening transition temperatures we use Eq.~(\ref{f03}) with a  correction of the growth kinetic coefficient by a factor of $\xi$ = 0.1, see~(\ref{f09}).
 The solid line is a calculation for the  first set of experiments and the dashed line is for the second experimental set. The increase of the crystal size shifts dependence $\delta$(T) towards higher temperatures.
\par
The curves Fast A,B are calculated by using the crystal growth rates in the abnormal state~\cite{VLT8} for the
 frequencies of the first and second runs, respectively.
  Crystals have the same size and mass but dependencies $\delta(T)$ at Fig.~\ref{fig1} differ by oscillation frequency.
  From the plot one can see that the transition
 of a crystal to the "burst-like growth" regime, which is accomplished by the drastic acceleration
 of the growth rate for all facets, results in a jump of decrement. The observation of such jump with keeping
  new magnitude of damping would evidence for a transition of the crystal to a steady anomalous state.

\section{EXPERIMENTAL RESULTS}
\par
The frequency and damping of the free oscillator in superfluid helium at low oscillation amplitudes are weakly
dependent on the temperature. The frequency increases by ~0.01\% along the melting curve for cooling
from 1.8K to 0.48K. The decrement is governed by the sound emission into the mounting base and lies
within the range of 0.002-0.003. As the oscillation amplitude increases to the velocity $\sim$1 m/s, the
damping enhances by one order of the magnitude. For small amplitudes, when the maximum rate does not exceed $\sim$10 cm/s, the decrement increases approximately linearly.
\par
\begin{figure}
\includegraphics[scale=1]{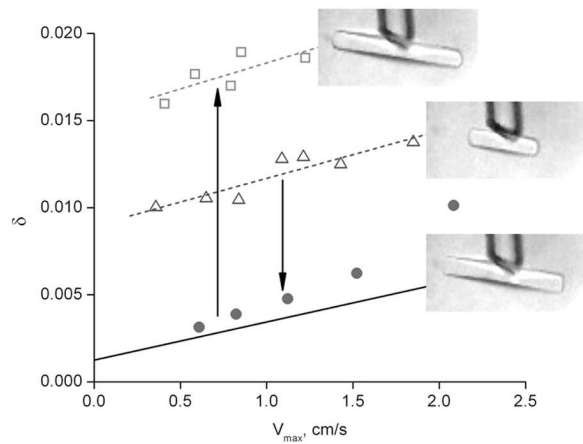}
\caption{The size effect of the atomically rough edges on the oscillator damping. The system damping: ($\bigtriangleup$) before growth, ($\bullet$) during growth, ($\square$) at the end of growth. The solid line is the free loop attenuation}
\label{fig3}
\end{figure}

The experiments are performed at 1.24K and 1.43K as well as below the first and second roughening
transitions (0.54K, 0.64K, 0.74K and 0.84K). Placing a crystal at the crossbar reduces the free oscillation
frequency of the oscillator in accordance with the model.
An additional damping contributed by the crystal above the roughening transition temperatures does not
exceed the natural damping of the oscillator in agreement with the calculation after Eq.(\ref{f09}).
The qualitative behavior of damping below the roughening transitions agrees well with the model proposed.
 In Fig.~\ref{fig3} the results are shown for the measurements of damping at the crystal before, during and after growth
  at 0.733K. Before starting the growth the crystal of diameter 2.4 mm and thickness 0.65 mm has large rounded
  edges giving an additional significant damping. With the beginning of the growth the pressure
  in the fluid increases, entailing the increase of the edge curvature and decrease of their area.
  As a result, the damping of the system drops to the magnitude close to the damping of the free loop.
  As the crystal growth stops, the pressure in the fluid relaxes to the phase equilibrium pressure,
  the size of the edges increases, and the attenuation intensifies. In Fig.~\ref{fig1} this evolution is shown by the squares
  in the small amplitude limit. The starting point is close to the theoretical curve for the first series of experiments.
  The drop of damping during the growth is shown by the downward arrow.
  The recovery to the higher magnitudes of the decrement, shown by the upward arrow,
  agrees also with the model proposed. The reason is that the increase of the crystal size shifts the
  dependence $\delta$(T) to the high temperatures region. On the right branch of the curve, this leads to the
  increased attenuation, as is observed in the experiment. The normal frequency of the system falls
  gradually from 442 Hz to 384 Hz with accumulating crystal mass in the growth process.
\par
\begin{figure}
\includegraphics[scale=1]{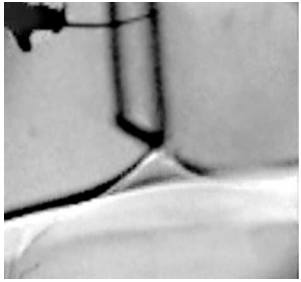}
\caption{The attachment of the atomically rough crystal surface to the oscillating crossbar. The temperature is 1.39 K, crossbar rate swing is 20 cm/s, and the spacing between the crossbar and undisturbed crystal surface is 0.7 mm.}
\label{fig4}
\end{figure}

The enhancement of the oscillation amplitude leads to the appearance of instability which results in sliding the
crystal down from the oscillator crossbar. The instability is observed in the both series of experiments at
a maximum oscillating rate more than $\sim$3~cm/s. For the first set of experiments in which the crystal
is pierced by the crossbar, the crystal starts to grow intensively until the crystal drops to the bottom of the cell.
 In the second run the crystal glides drastically down as the critical rate is reached. We have attempted to put
  the crystal from the needle to the crossbar oscillating with the large amplitude, expecting that the crystal
  could be attracted to the oscillator as it was observed for the oscillations of the crossbar beside the surface,
  see Fig.~\ref{fig4}. However, this produced no result. For the oscillating rates with a maximum velocity above
  $\sim$3~cm/s, the crystal bounces from the crossbar. The existence of such instability limits the maximum
  magnitude of the overpressure by $\sim$0.2 mbar at the critical point of the streamline of the crystal.
  This value is one order of the magnitude less than that required for the transition of the crystal into an abnormal state\cite{VLT9}
  \par
   Below $T_{R1,R2}$ in the steady state the crystal is suspended at the needle tip for hours\cite{VLT6}. The Rayleigh-Taylor instability does not develop at the bottom facet during this time as a result of a giant potential barrier which should be overcome via either thermal activation\cite{BDT2} or quantum tunneling\cite{BD}. For this reason, the instability observed in our experiment is associated directly with a motion of the crystal.
  \par
 In the reference frame associated with the oscillator crossbar at the acceleration maximum one half of the crystal is under conditions of appearing the Rayleigh-Taylor instability. However, the acceleration magnitude is about $\sim 10^4 cm/s^2 \sim10g$ at the velocity $\sim 3 cm/s$ and frequency $\sim 500 Hz$, $g$ being the acceleration of gravity. This is still insufficient for overcoming the potential barrier during the oscillation half-period\cite{BDT2,BD}. An accelerating motion can give rise to the Richtmayer-Meshkov instability\cite{BDT3}. However, the estimates show that the acceleration magnitude is too small to develop this instability.
 \par
   The fluid flow along the facet, as is shown by Andreev\cite{A}, decreases the energy of the steps normal to the fluid flow and increases its concentration. These processes, in principle, may intensify the classical mechanisms of the facet growth. To conclude, the origin of the instability observed as sliding the crystal from the oscillator crossbar is not established.

\section{SUMMARY}
\par
The kinetics of atomically rough surfaces in a faceted $^4$He crystal affects significantly the motion
of the crystal in a superfluid liquid and governs the hydrodynamics of the streamline.
The damping of the oscillating crystal is determined mainly by the temperature behavior
of the kinetic growth coefficient and an area of rough edges.
The model proposed describes well the qualitative features of the phenomenon and satisfactory
the numerical magnitudes of the damping. Unfortunately, the appearance
of hydrodynamic instability for the oscillations at a rate higher than 3~cm/s prevents us
from obtaining high overpressures at the crystal surface. Hence, the questions whether
it is possible to prepare the crystal in the stationary abnormal state by oscillating the crystal has remained open.

\section{ACKNOWLEDGMENTS}
This work was supported by the RFBR grant No.08-02-00752-a..

\end{document}